\documentclass[prl, aps, twocolumn, hyperref, superscriptaddress]{revtex4}

\usepackage{graphicx}

\usepackage{amssymb, amsmath, color, rotating}
\begin{document}

\title{Striped Ferronematic ground states in a spin-orbit coupled $S=1$ Bose gas}

\author{Stefan S. Natu}
\email{snatu@umd.edu}
\affiliation{Condensed Matter Theory Center and Joint Quantum Institute, Department of Physics, University of Maryland, College Park, Maryland 20742-4111 USA}

\author{Xiaopeng Li}
\affiliation{Condensed Matter Theory Center and Joint Quantum Institute, Department of Physics, University of Maryland, College Park, Maryland 20742-4111 USA}

\author{William S. Cole}
\affiliation{Condensed Matter Theory Center and Joint Quantum Institute, Department of Physics, University of Maryland, College Park, Maryland 20742-4111 USA}

\begin{abstract}
We theoretically establish the mean-field phase diagram of a homogeneous spin-$1$, spin-orbit coupled Bose gas as a function of the spin-dependent interaction parameter, the Raman coupling strength and the quadratic Zeeman shift. We find that the interplay between spin-orbit coupling and spin-dependent interactions leads to the occurrence of ferromagnetic or ferronematic phases which also break translational symmetry. For weak Raman coupling, increasing attractive spin-dependent interactions 
(as in $^{87}$Rb or $^7$Li) induces a transition from a uniform to a stripe XY ferromagnet (with no nematic order). For repulsive spin-dependent interactions however (as in $^{23}$Na), we find a transition from an $XY$ spin spiral phase ($\langle S_{z} \rangle = 0$ and uniform total density) with uniaxial nematic order, to a biaxial ferronematic, where the total density, spin vector and nematic director oscillate in real space. We investigate the stability of these phases against the quadratic Zeeman effect, which generally tends to favor uniform phases with either ferromagnetic or nematic order but not both. We discuss the relevance of our results to ongoing experiments on spin-orbit coupled, spinor Bose gases.
\end{abstract}
\maketitle

\section{Introduction} The interplay between competing orders such as superfluidity/superconductivity, magnetism, liquid crystallinity and density wave order is fundamental to the rich phenomenology of strongly correlated systems. A candidate system for exploring this physics is a spin-orbit coupled Bose condensate \cite{galitski, nistexpt, zhang1, ji}, where the coupling between spin and motional degrees of freedom can lead to a spin textured ground state which breaks rotational symmetry in spin space \cite{stanescu, congjun, wang}, as well as translational symmetry in real space \cite{jasonspinorbit, stringari}. Indeed, for a pseudospin-$1/2$ Bose system, varying the spin-orbit coupling strength drives a transition from an unmagnetized phase with density wave order to a uniform magnetized phase, which has been studied both theoretically and experimentally \cite{stringari, ji}. More recently, attention has turned towards exploring the physics of large spin systems, which have no analog in condensed matter, such as highly magnetic atoms like Dysprosium, Erbium, Chromium \cite{pfauchromium, erbiumnew, levdyraman}, and alkaline earth atoms with SU(N) symmetry (See Ref.~\cite{cazalilla} and referenes therein). The large spin nature of these atoms produces a rich phase diagram with novel topological defects, where uniaxial and biaxial nematic and more exotic platonic solid orders compete and complement conventional magnetically ordered phases \cite{dienerho, huangho, muellerrotating}. In the presence of spin-orbit coupling, the possibility of translational symmetry breaking can lead to textured ground states phases with intertwining magnetic and nematic order. Here we study the simplest, experimentally realizable system where such physics is manifest: a spin-$1$, spin-orbit coupled Bose gas \cite{ohberg, wang}, finding a rich phase diagram. 

\begin{figure}
\begin{picture}(200, 180)
\put(-15, -10){\includegraphics[scale=0.55]{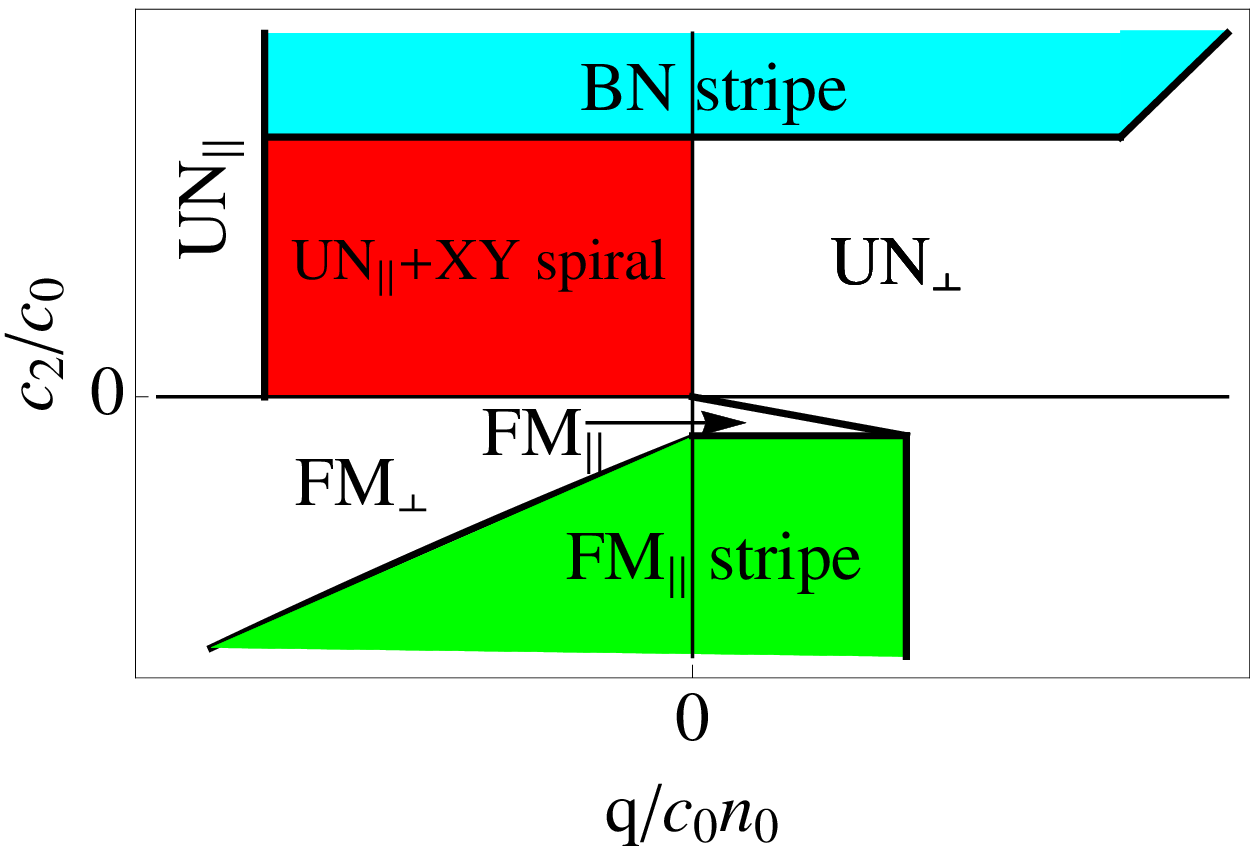}}
\put(58, 125){\includegraphics[scale=0.2]{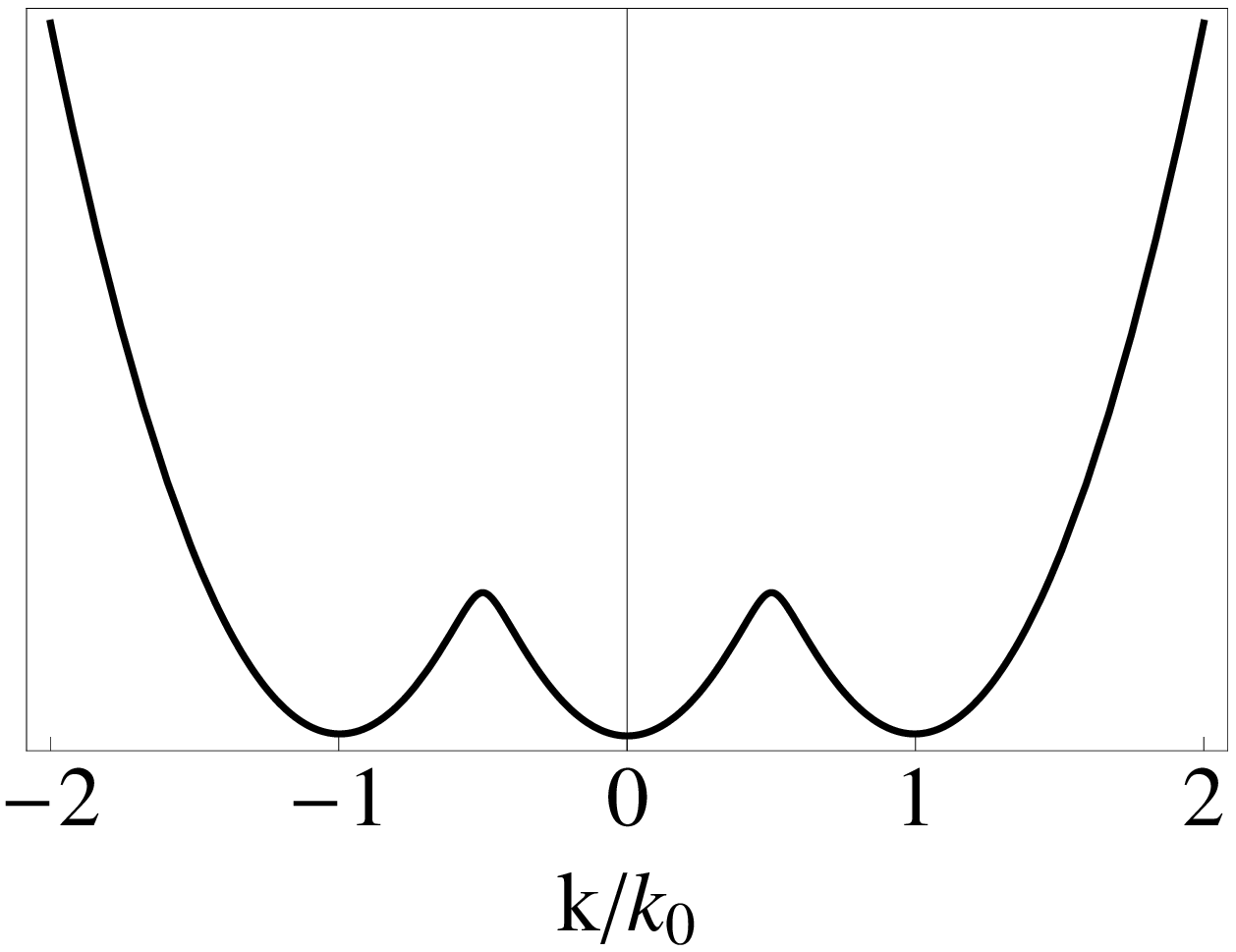}}
\put(-23, 131){\includegraphics[scale=0.21]{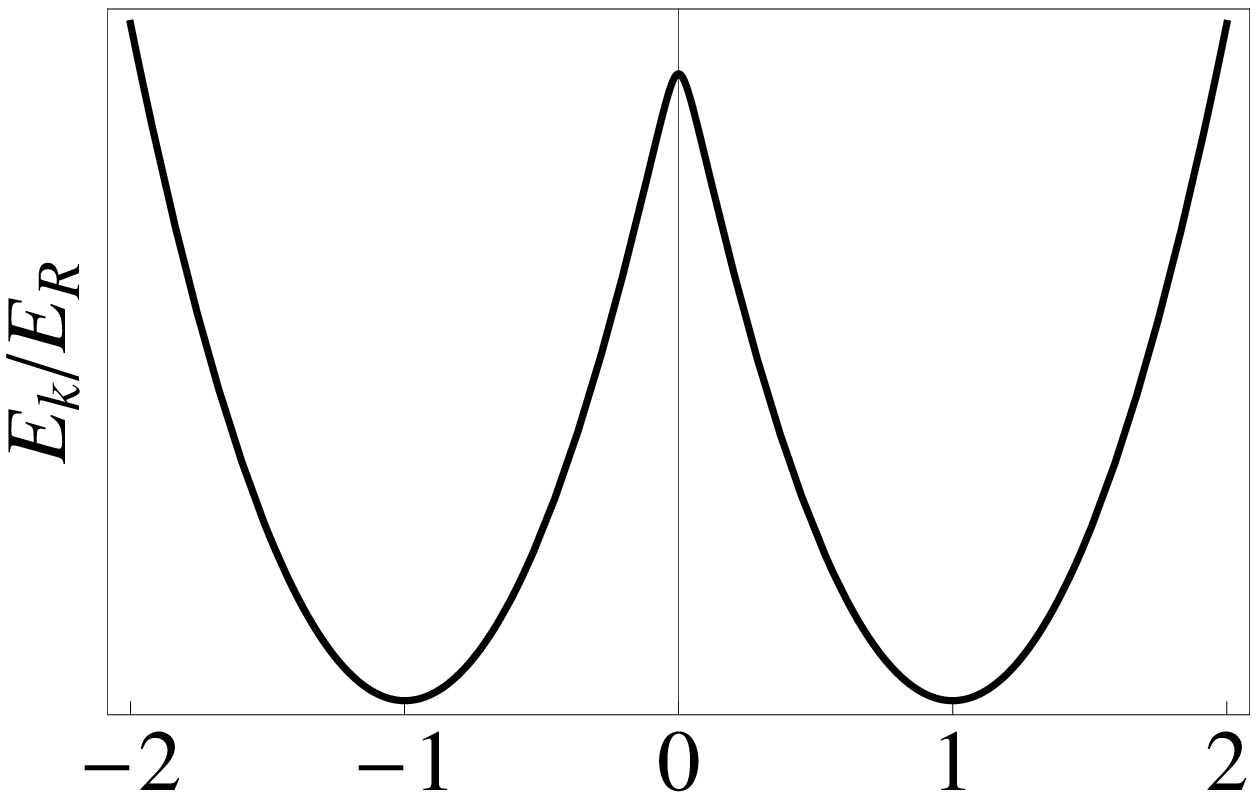}}
\put(135, 131){\includegraphics[scale=0.2]{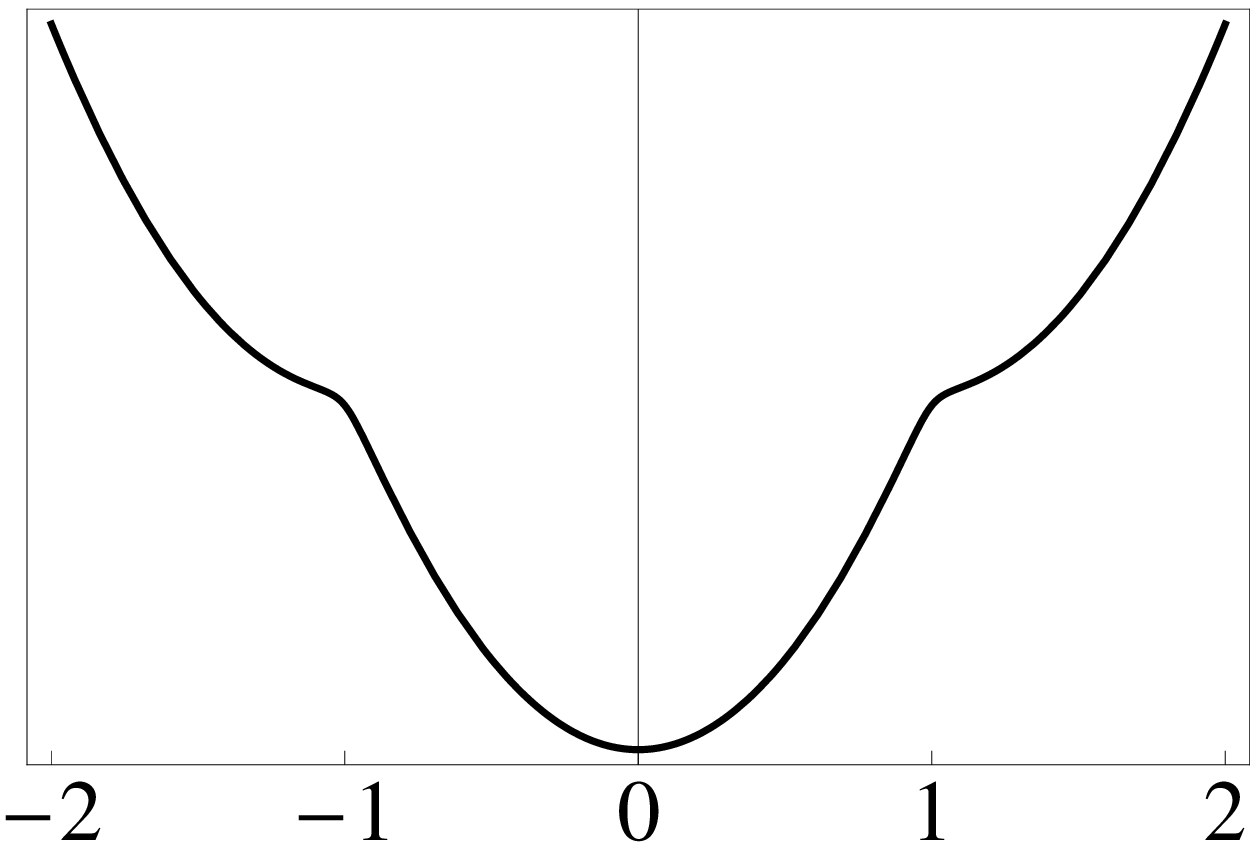}}
\end{picture}
\caption{\label{pd} (Color Online) Schematic phase diagram for a spin-orbit coupled spin-$1$ Bose gas, as a function of spin-dependent interaction strength and quadratic Zeeman shift, showing translation symmetry breaking phases. The underlying single-particle dispersion is shown above. For sufficiently large attractive spin-dependent interactions, an XY spin density wave phase occurs, with oscillations in the total density ($FM$ stripe). For repulsive spin-dependent interactions, an XY spiral phase occurs, which simultaneously has uniaxial nematic order (UN), but the total density remains uniform.  For sufficiently large spin-dependent interactions, a biaxial nematic phase ($BN$) is present where the total density, spin vector and nematic director oscillate in real space. Sufficiently large positive or negative quadratic Zeeman effect favors homogeneous phases with either uniaxial nematic ($UN_{\perp}$ ($\psi = \{0, 1, 0\}$) or $UN_{\parallel}$ ($\psi = \frac{1}{\sqrt{2}}\{1, 0, 1\}$)) or ferromagnetic order ($FM_{\perp}$ ($\psi = \{1, 0, 0\}$) or $FM_{\parallel}$ ($\psi = \frac{1}{2}\{1, \sqrt{2}, 1\}$)), but not both.}
\end{figure}

Our main result is summarized in Fig.~\ref{pd}, which shows the schematic, zero-temperature phase diagram of a spin-orbit coupled spin-$1$ Bose gas as a function of the spin-dependent interaction and the quadratic Zeeman energy, at fixed Raman coupling and spin-independent interaction strength. A new feature of the spin-orbit coupled gas is the appearance of translational symmetry breaking phases with simultaneous spin and nematic order, which are generically competing orders in this system \cite{muellerrotating, dienerho}. Weak repulsive spin-dependent interactions favor a uniaxial nematic ferromagnet (ferronematic), which supports a spiral spin texture in the $x-y$ plane ($UN + XY$ spiral). Large attractive spin-dependent interactions favor a ferromagnetic stripe ($FM$ stripe), whereas large repulsive spin-dependent interactions favor a \textit{biaxial} ferronematic stripe phase ($BN$ stripe) where the total density, spin vector and nematic director oscillates in real space. 

\section{Spin-$1$ Phenomenology} In the absence of spin-orbit coupling, the phase diagram of a spin-$1$ Bose gas has been well established theoretically \cite{ho, ohmi, law, mukherjee1} and experimentally \cite{stenger, chang, stamper-kurn, stamperkurnreview, sadler, jacob}. Assuming short-range (s-wave) interactions, spin rotation invariance and bosonic statistics forces two-body collisions to occur in the total spin-$0$ or spin-$2$ channels, producing the interaction Hamiltonian \cite{ho, ohmi}: 
\begin{equation}\label{intham}
{\cal{H}}_{int} = \frac{1}{2}\int d^{3}\textbf{r}~\psi^{\dagger}_{\alpha}\psi_{\beta}^{\dagger}\psi_{\gamma}\psi_{\delta}(c_{0}\delta_{\alpha\delta}\delta_{\beta\gamma} + c_{2} \textbf{S}_{\alpha\delta}\cdotp \textbf{S}_{\beta\gamma}), 
\end{equation}
where the greek indices denote the hyperfine spin projection, and $\psi_{\sigma}(r)$ is the the boson field operator. Unlike the pseudospin-$1/2$ case, this Hamiltonian has SU$(2)$ spin rotation invariance. 

The two coupling constants, $c_0$ and $c_2$ represent spin-independent and spin-dependent interactions respectively, and $\textbf{S}$ is the vector $\{S^{x}, S^{y}, S^{z}\}$, where $S^i$ are $3 \times 3$, spin-1 matrices. The interactions are expressed in terms of the microscopic scattering lengths in the spin-$0$ ($a_{0}$) and spin-$2$ ($a_{2}$) channels and atomic mass $m$ as: $c_{0} = 4\pi (a_{0}+ 2a_{2})/3m$ and $c_{2} = 4\pi(a_{2} -a_{0})/3m$
\cite{stamperkurnreview}. 
For $^{87}$Rb, $c_{2}/c_{0} = -0.005$, for $^{23}$Na, $c_{2}/c_{0} \sim 0.05$ and for $^{7}$Li, $c_{2}/c_{0} \sim -0.5$ \cite{stamperkurnreview}. These interactions (and their sign) can however be tuned using optical Feshbach resonances \cite{fatemi}. 

The wave-function of a spin-$1$ Bose condensate is represented as a spinor $\psi = e^{i\theta}\{\psi_{1}, \psi_{0}, \psi_{-1}\}$, where $\theta$ represents the broken global $U(1)$ gauge symmetry of the Bose condensate, and $\{1, 0, -1\}$ label the three spin states. Owing to the structure of the spin-$1$ Pauli matrices, this system can exhibit both magnetic order, given by the vector order parameter $\langle \textbf{S} \rangle = \langle \psi|\textbf{S}|\psi\rangle/n$, where $n$ is the density, or nematic order, described by the tensor ${\cal{N}}_{\mu\nu} = \delta_{\mu\nu} - \frac{1}{2n}\langle \psi|(S_{\mu}S_{\nu}+S_{\nu}S_{\mu})|\psi\rangle$, where $\{\mu, \nu\} \in \{x, y, z\}$, and $\delta_{\mu\nu}$ denotes the identity matrix. However, as pointed out by Mueller \cite{muellerrotating}, these orders are not independent of one another, but rather competing. Diagonalizing the nematic tensor yields three distinct eigenvalues $\lambda_{1}, \lambda_{2}, \lambda_{3}$, constrained by $\lambda_{1}+\lambda_{2}+\lambda_{3} = 1$. A \textit{uniaxial} nematic has one non-zero eigenvalue, while a \textit{biaxial} nematic has three distinct eigenvalues. 
Attractive spin-dependent interactions  ($c_{2} < 0$) favor a maximally ferromagnetic phase ($\langle \psi|\textbf{S}|\psi\rangle =\hat{\textbf{z}}$), with no nematic order, which can be represented by unitary rotations of $\psi = e^{i\theta}{\cal{U}}\{1, 0, 0\}^{T}$, whereas repulsive spin-dependent interactions ($c_{2}>0$) favor a uniaxial nematic with no spin order, represented by unitary rotations of $\psi = e^{i\theta}{\cal{U}}\{0, 1, 0\}^{T}$ \cite{ho}. 

In the presence of spin-orbit coupling, spin-rotation symmetry is broken and the three spin states are no longer degenerate at the single-particle level. We choose the spin-orbit coupling to be of equal Rashba-Dresselhaus type, which was recently realized in experiments \cite{nistexpt, zhang1, chinasoc, zwierleinsoc, spielman2, ji}, but generalized to the spin-$1$ case:
\begin{equation}\label{soham}
{\cal{H}}_{soc} = \frac{\hbar^{2}(k_{x} - k_{0}S_{z})^{2}}{2m} + \frac{\hbar^{2}k^{2}_{\perp}}{2m} + \frac{\Omega}{2}S_{x} + \frac{\delta}{2}S_{z} + \frac{q}{2}S^{2}_{z}
\end{equation}
where $k_{0}$ is the wave-vector of the Raman beams, $\Omega$ is the strength of the Raman coupling, $\delta$ and $q$ are the linear and quadratic Zeeman effects respectively. It is convenient to normalize energy by the recoil energy of the Raman lasers $E_{R} = \frac{\hbar^{2}k^{2}_{0}}{2m}$. For simplicity, we neglect the linear Zeeman effect term ($\delta = 0$) in this work, but generally assume that $q \neq 0$, and can take on positive and negative values, which can be achieved using microwaves \cite{vinit, gerbierqtune}.

A detailed description of the single-particle physics of a spin-$1$ spin-orbit coupled gas was recently given by Lan and \"Ohberg \cite{ohberg}, and is not repeated here. 
For weak $q$ and $\Omega$, the low energy spectrum has three local minima at $k = 0, \pm k_{1}$, where $0 \leq k_{1}/k_{0} \leq 1$ is determined by diagonalizing Eq.~\ref{soham}, at fixed $\Omega$ and $q$. Increasing positive $q$ results in a single minimum at $k=0$, whereas negative $q$ produces a two minimum structure, with the minima at $\pm k_{1}$. 

The dispersion of the lowest branch (to quadratic order in $k_x$) is obtained as 
\begin{equation}
\epsilon (k_x) = \frac{\hbar^2 k_x^2}{m} \left[1/2-\frac{\hbar^2 k_0^2 }{m} (\tilde{q} ^2 + 4 \Omega^2)^{-1/2} \frac{1-z}{1+z}  \right] 
+ {\cal O} (k_x^4), 
\end{equation}
with $\tilde{q} = q+ \frac{\hbar^2 k_0^2}{m}$ and 
$z = \frac{\tilde{q}}{\sqrt{\tilde{q}^2 + 4 \Omega^2}}$. 
The analytic form of the transition line from three to two minima easily follows as 
$
\frac{\hbar^2 k_0^2 }{m} (\tilde{q} ^2 + 4 \Omega^2)^{-1/2} \frac{1-z}{1+z} = 1/2. 
$

From the triple minimum structure of single-particle dispersion, 
we make a reasonable variational ansatz 
for the condensate wave-function:
\begin{equation}\label{wavefn}
\psi = \sqrt{\frac{N}{V}}(\chi_{+}e^{ik_{1}x}\phi_{+} + \chi_{0}\phi_{0} + \chi_{-}e^{-ik_{1}x}\phi_{-}),
\end{equation}
where $\chi_{0}, \chi_{\pm}$ are complex numbers, which are determined variationally below, $N$ is the particle number, $V$ is the volume, and $\phi_{\pm}, \phi_{0}$ are the normalized single-particle spinor eigenstates at the minima $\pm k_{1}, 0$ respectively. We fix the gauge choice by choosing  the eigenvectors ($\phi_{\pm}, \phi_{0}$) to be real, where the respective spin components obey $\phi^{\pm 1}_{\pm} = \phi^{\mp 1}_{\mp}$, and $\phi^{0}_{+} = \phi^{0}_{-}$. Particle number conservation $N = \int d^{3}\textbf{r}~n(\textbf{r}) = \sum_{\sigma \in \{-1, 0, 1\}}\int d^{3}\textbf{r} |\psi_{\sigma}(\textbf{r})|^{2}$ imposes the constraint $|\chi_{+}|^{2}+|\chi_{0}|^{2}+|\chi_{-}|^{2} =1$. 

The variational interaction energy takes a suggestive form: 
\begin{equation}\label{energy} 
E = r (|\chi_+| ^2 + |\chi_-|^2 ) + g_{\mu \nu} |\chi_\mu |^2|\chi_\nu|^2 + g_3 (\chi_+ ^* \chi_-^* \chi_0 \chi_0 + c.c.) , 
\end{equation}
where $r$ is the kinetic term and $g_{\mu\nu}$ and $g_{3}$ are related to the original interaction parameters multiplied by form factors proportional to the single-particle wave-functions at the three minima. The energy is invariant under transformations $U_C (1): \chi_\mu  \to e^{i\theta} \chi_\mu$, $U_A(1): \chi_\mu \to e^{ i\mu \theta} \chi_\mu$ 
and $Z_2: \chi_\mu \to \chi_{-\mu}$. Here the axial $U_A(1)$ originates from translational symmetry, and the $Z_2$ symmetry is related to reflection where spin transforms as a pseudovector. It is important to emphasize that the Josephson term, proportional to $g_{3}$ is zero throughout the phase diagram of the spin-$1$ Bose gas without spin-orbit coupling. However, as we show here, it plays a crucial role in the physics of the spin-orbit coupled gas.  

A new feature of the spin-orbit coupled Bose gas is the possibility of translational symmetry breaking phases \cite{wang, jasonspinorbit}, which arise because bosons in different spin states condense into finite momentum states. For the spin-$1/2$ case, where bosons condense at two minima, the total density develops stripes at a wave-vector $2k_{1}$ \cite{wang, jasonspinorbit, stringari}, but the spin density $\langle \textbf{S} \rangle$ remains uniform throughout the phase diagram \cite{note, stringari}. In the spin-$1$ case however, in addition to stripe structure in the density \cite{wang, ohberg}, the system can display oscillations in the spin and nematic order parameters. This leads to a rich phase diagram, reproduced in Fig.~\ref{pd}, which we now discuss in detail. In Table \ref{orders}, we characterize the condensed phases we find, in terms of their order parameters. 

\begin{table}[tb]
\caption{Orders in spin-orbit coupled spin-1 gas.}\label{orders}
\begin{ruledtabular}
\begin{tabular}{lcr}
\bf Order&\bf Symbol&\bf Order Parameter\\
ferromagnetic & $FM_{\parallel/\perp}$&$\langle S^i \rangle \neq 0$\\
Uniaxial nematic&$UN_{\parallel/\perp}$&$\lambda_{1} \neq 0, \lambda_{2} = \lambda_{3} =0$\\
Biaxial nematic&$BN$&$\lambda_{1} < \lambda_{2} < \lambda_{3}$\\
Translation& stripe, $XY$ spiral &$\langle S^i(\textbf{r}) \rangle \sim \cos(k_{1}r)$\\
& & $n(r) \sim \cos(k_{2} r)$\\
\end{tabular}
\end{ruledtabular}
\end{table}

We minimize the total energy per particle (Eq.~\ref{energy}), with respect to the complex variational parameters $\chi_{0}, \chi_{\pm}$. The spin and nematic order parameters are then computed using the resulting mean-field ground state wave-function. Normalizing the energy to the laser recoil energy $E_{R}$, and setting $\delta =0$, yields four dimensionless parameters $\Omega/E_{R}$, $q/E_{R}$, $c_{0}n_{0}/E_{R}$, $c_{2}n_{0}/E_{R}$, where $n_{0} = N/V$ is the total density. Throughout, we fix the Raman coupling $\Omega/E_{R}$ such that, in the absence of a quadratic Zeeman effect, the low energy, single-particle dispersion has three local minima. The two minima at $k = \pm k_{1}$ are always degenerate in the absence of $\delta$. For $q>q_{c1} > 0$ the three minimum structure disappears and only a single minimum at $k =0$ is present, whereas for $q < q_{c2} <0$, the system only has two minima at finite $k$. We fix $\Omega$ but vary $q$ to access both these regimes in parameter space \cite{ohberg}.   

\section{Attractive spin-dependent interactions $c_{2} < 0$} We first consider the regime of attractive spin-dependent interactions, which corresponds to $^{87}$Rb (as in the NIST experiments \cite{nistexpt, spielman2}) and $^{7}$Li. Absent spin-orbit coupling, the ground state is a uniform ferromagnet, which is of Ising type ($\langle \textbf{S} \rangle =\hat{\textbf{z}}$) for $q<0$ and XY type ($\langle \textbf{S} \rangle = \hat{\textbf{x}}$) for $q>0$ (spin rotation symmetry is restored at $q=0$). For sufficiently large $q>0$, there is a second order quantum phase transition to a polar ($UN_{\perp}$) phase, which has been investigated in detail recently (see Ref.~\cite{stamperkurnreview} and References therein). 

In the presence of spin-orbit coupling, at $q=0$, spin symmetry is explicitly broken by the Rabi coupling ($\Omega$) term, which prefers to align the total magnetization along $x$. Absent interactions, the single particle wave function is centered around $k=0$, and has a small but finite value of $\langle S_{x} \rangle$, corresponding to the explicitly broken symmetry. Upon turning on $c_{2}$, the minimum energy state is a ferromagnet with $\langle \textbf{S} \rangle = \hat{\textbf{x}}$. The wave-function for such a state requires all three minima to be occupied, but the relative phase $\theta_{+} + \theta_{-} - 2 \theta_{0} \approx 0$, where $\chi_{\pm 0} = |\chi_{\pm, 0}|e^{i\theta_{\pm,0}}$. This state is degenerate with the plane wave Ising ferromagnet formed by occupying a single minimum at $k = \pm k_{1}$ with respect to the spin-dependent interaction term, but has stripes in the total density, and is thus penalized by the repulsive density-density interactions (proportional to $c_{0}$). For weak $|c_{2}|$ therefore, an Ising plane wave phase occurs, which has no stripes in the total density. Upon increasing $|c_{2}|$ however, the total energy can be lowered by aligning the spin along the $x$ direction, satisfying the Rabi coupling term, at the expense of producing density (and spin) stripes. This interaction driven transition from a uniform Ising ferromagnet to striped XY ferromagnet is a new feature of the spin-$1$, spin-orbit coupled gas \cite{ohberg}. 



\begin{figure}
\begin{picture}(200, 95)
\put(0, -10){\includegraphics[scale=0.43]{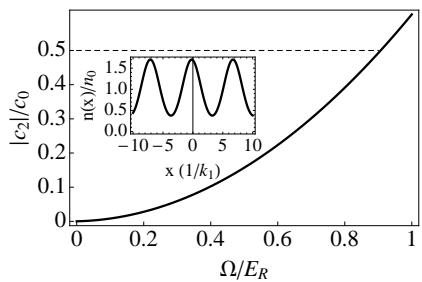}}
\end{picture}
\caption{\label{ferrostripeonset} Critical attractive spin-dependent interaction $|c_{2}|/c_{0}$ for onset of stripe ferromagnetic phase as a function of Rabi coupling $\Omega/E_{R}$ at $q=0$. We set $c_{0}n_{0}/E_{R} = 0.4$. Dashed line shows $|c_{2}|/c_{0}$ for $^{7}$Li; all values of $\Omega \lesssim 0.9 E_R$ support the ferromagnetic phase with stripes in the total density. Inset shows the density in real space for $\Omega/E_{R} = 0.8$ at $|c_{2}|/c_{0}$ corresponding to $^{7}$Li.} 
\end{figure}

In Fig.~\ref{ferrostripeonset} we plot the critical value of $|c_{2}|/c_0$ for the ferromagnetic stripe phase as a function of the Rabi strength $\Omega/E_{R}$. For our parameters, when $\Omega/E_{R} > 1$, the system enters the single minimum regime, and the stripe phase is destroyed. Although the stripe phase can occur for arbitrarily small spin dependent interactions, it should be emphasized that the amplitude of the stripes decreases with decreasing $|c_{2}|/c_{0}$, and may be hard to resolve experimentally, particularly for $^{87}$Rb. The horizontal dashed line shows the spin-dependent interaction for $^{7}$Li \cite{stamperkurnreview}; all values of Rabi coupling below the line should exhibit a stripe phase. The inset shows the pronounced amplitude of the stripes for the $^{7}$Li interaction parameters at $\Omega/E_{R} = 0.8$, which strongly supports the experimental observability of this phase. 

As shown in Fig.~\ref{pd}, the striped ferromagnetic phase is destroyed for positive and negative values of the quadratic Zeeman effect. For $q>0$, we find a transition from the stripe ferromagnetic phase to a polar condensate, which occurs roughly where the single-particle dispersion enters the single minimum regime ($q_{c1}$), largely independent of $c_{2}$. For $q<0$, the single-particle dispersion has two minima, and a uniform Ising ferromagnetic phase occurs where only one of these two degenerate minima are occupied. The transition from the stripe ferromagnet to the uniform Ising ferromagnet where translation symmetry is restored, depends on the interaction strength and the magnitude of $q$ as shown in Fig.~\ref{pd}.


\section{Polar Regime: $c_{2} > 0$} We now turn our attention to the case of repulsive spin-dependent interactions. Absent spin-orbit coupling, repulsive spin-dependent interactions yield a polar phase, where $\langle \textbf{S} \rangle = 0$, but the system has uniaxial nematic order. In the presence of the Rabi term $\Omega$, residual ferromagnetic order is present even for $c_{2} > 0$, and generically, the ground state is ferronematic. 

 For weak spin-dependent interactions and $q>0$, the system condenses at $k =0$, and a uniform ferronematic phase is found. For $q<0$ however, the system condenses at $k =\pm k_{1}$, and the phases of the condensate at these two points are such that the total density remains uniform, but the transverse spin density spontaneously breaks translational symmetry and develops XY spin density wave order. Similar spiral phases have been predicted to occur in the spin-$1/2$ system in the presence of Rashba spin-orbit coupling \cite{congjun, wang}. 

This origin of the spin density wave can also be understood as follows: For weak Raman coupling, the condensates at $k = \pm k_{1}, 0$ are closely related to the original $\pm 1, 0$ spin states. Thus by applying a gauge transformation, whereby $\psi_{\pm 1} \rightarrow \psi_{\pm 1}e^{\mp i k_{1}x}, \psi_{0} \rightarrow \psi_{0}$, we obtain a non spin-orbit coupled Hamiltonian with a spatially oscillating magnetic field along $x$ of the form $\Omega S_{x} \cos(k_{1}x)$, leaving all other terms in the Hamiltonian unchanged. This produces an XY spin density wave texture in real space. As $q$ becomes more and more negative, the amplitude of oscillation of the spiral goes to zero as $\Omega/|q|$. The nematic tensor in this phase has only one non-zero eigenvalue, which corresponds to a uniaxial nematic. Thus spin-orbit coupling naturally leads to ferronematic phases which break translation symmetry.

\begin{figure}
\begin{picture}(200, 140)
\put(0, 110){\includegraphics[scale=0.45]{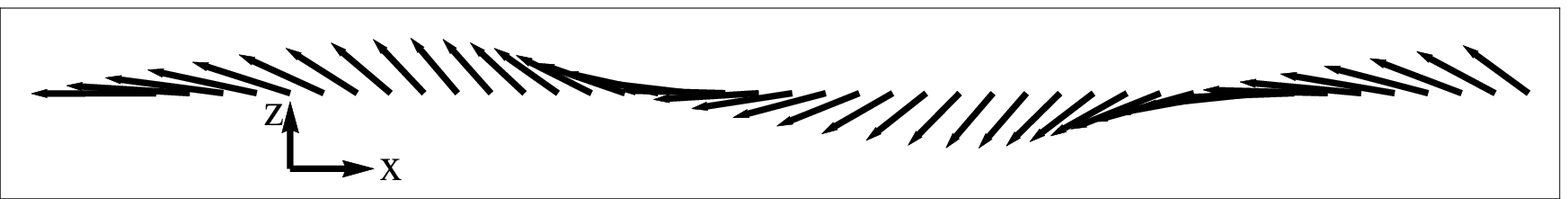}}
\put(0, -10){\includegraphics[scale=0.45]{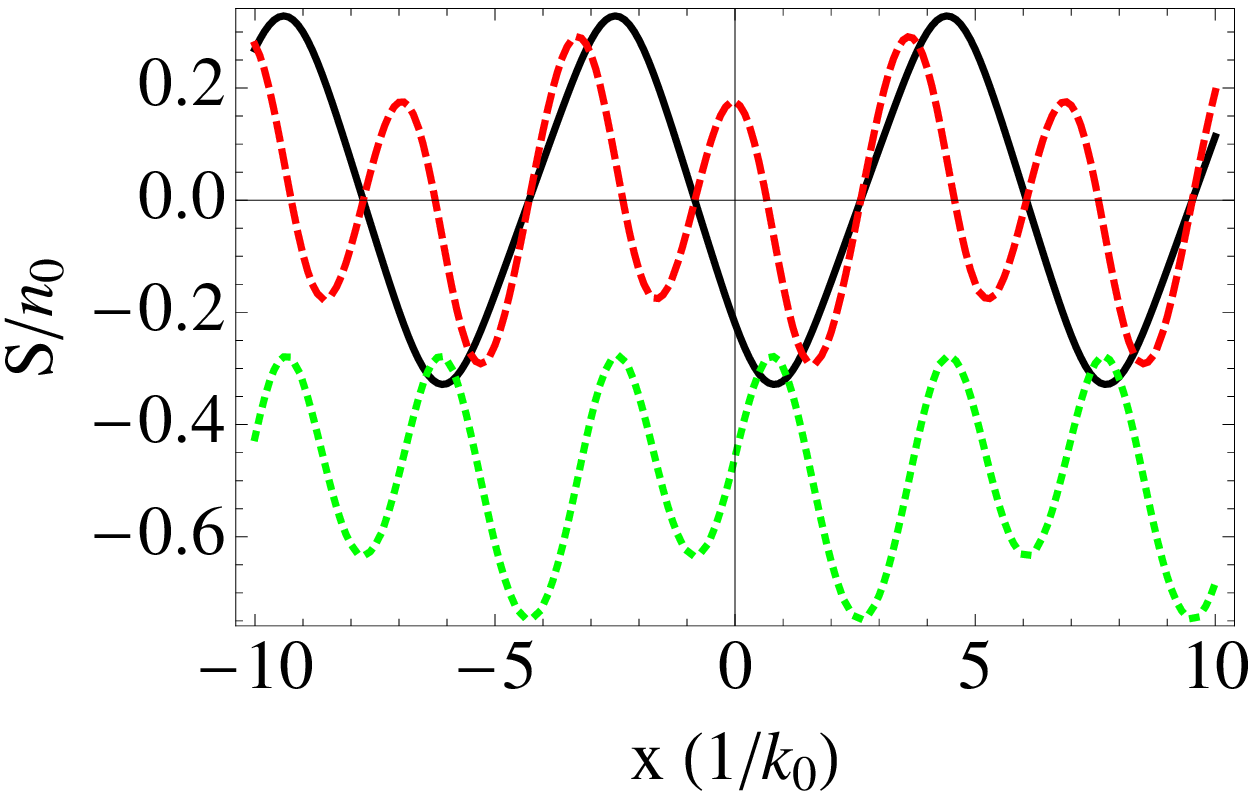}}
\end{picture}
\caption{\label{spinfig} Top: Spatial spin texture in the polar regime of a spin-orbit coupled spin-$1$ Bose gas for strong spin-dependent interactions $c_{2}/c_{0} = 1.1$, and $\Omega/E_{R} = 0.8$, which corresponds to the three minimum regime. We set $q = 0$ here, although this phase is stable for moderate values of $q$ (see Fig.~\ref{pd}). The arrows indicate the projection of the spin in the $x-z$ plane in spin space at each point in real space. Bottom: $x$ (green dotted), $y$ (red dashed) and $z$ (black solid) components of spin in real space. Oscillations in the $z$ direction (at wave-vector $k = k_{1}$) and $x-y$ direction (at two wave-vectors $k = k_{1}, 2k_{1}$) are different with one another. The total density (not shown) also shows anharmonic oscillations at two dominant wave-vectors $k = k_{1}$ and $k = 2k_{1}$. Note that due to the Raman coupling term, time reversal is explicitly broken, and the spatially averaged spin along the $x$ direction is finite.} 
\end{figure}

For even larger $c_{2}$, the situation becomes more exotic, and the system condenses into all three minima, even at $q=0$. The relative phase of the condensates at the three minima $\theta_{+} + \theta_{-} - 2\theta_{0} \approx \pi$. This state has stripes in the total density \cite{ohberg}, and concurrently, magnetic order in all $x, y,z$ directions, as shown in Fig.~\ref{spinfig}. Interestingly however, the wavelengths of the oscillations in $z$ and those in the $x-y$ directions are generally different from one another.  

\begin{figure}
\begin{picture}(200, 100)
\put(0, -10){\includegraphics[scale=0.45]{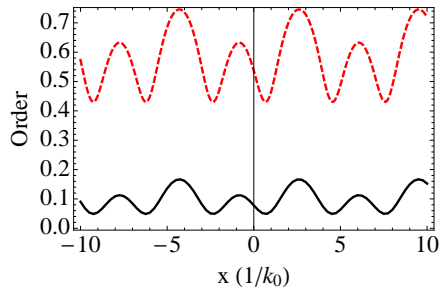}}
\end{picture}
\caption{\label{biaxialfm} Total spin (red dashed) and biaxial order parameter (black, solid) shown as a function of space in the biaxial, spin-density wave phase of a polar spin-orbit coupled gas. The minimum in the biaxiality corresponds to a minimum in the total spin, and here a nearly uniaxial nematic appears, which illustrates the competing nature of the nematic and spin order parameters.} 
\end{figure}

Furthermore, diagonalizing the nematic tensor for this situation yields three distinct eigenvalues, which is a \textit{biaxial} nematic phase. Owing to the constraint $\lambda_{1}+\lambda_{2}+\lambda_{3} =1$, the degree of \textit{biaxiality} can be found by taking the difference between the largest two eigenvalues. As the density, spin and nematic order parameters are not independent of one another, all these orders simultaneously oscillate in real space. In Fig.~\ref{biaxialfm}, we plot the spatial oscillations in the biaxiality and the total spin,  the minimum in the biaxiality coincides with the minimum in the total spin, where a maximally uniaxial nematic phase occurs. 

\section{Experimental Relevance} As we show in Fig.~\ref{ferrostripeonset}, observing the ferromagnetic stripe phase for attractive spin-dependent interactions may be challenging for $^{87}$Rb, but feasible in $^{7}$Li. Although the contrast in the oscillations in the spin density is large, experiments may not have enough spatial resolution to observe the individual oscillations, which will further be smeared out by spatial averaging effects during expansion and imaging. Martone \textit{et al.} \cite{martone} have recently proposed using Bragg spectroscopy to increase the wavelength of the stripes, making them visible. 

The XY spiral nematic state occurs even for weak, repulsive spin-dependent interactions (as in $^{23}$Na spinor condensates) by simply tuning the quadratic Zeeman shift to take on negative values, which produces two global minima in the single-particle dispersion. The transverse components of spin can then be probed \textit{in situ} by applying spin echo radio-frequency pulses between individual snapshots  \cite{guzman} to reveal the spin density texture. Nematicity can be probed experimentally by directly measuring spin fluctuations $\langle S_\mu S_{\nu} \rangle$, averaged over many shots \cite{hamley}. Alternatively, optical birefringence, whereby, the coupling between the nematic order parameter and the polarization of a probe beam leads to a rotation in the polarization of the light field can be used to measure the local nematic/biaxial order parameter \cite{carusotto}. Observing the biaxial spin-density wave phase requires strong spin-dependent interactions, which could be induced using optical or magnetic Feshbach resonances. This however breaks the SU$(2)$ spin rotation invariance which underpins Eq.~\ref{intham}, and may lead to a qualitatively different phase diagram than that discussed here.  

\section{Conclusions and Future Directions} In conclusion, we find that the spin-$1$ spin-orbit coupled Bose gas possesses a rich phase diagram with phases which break translational symmetry, spin symmetry and possess liquid crystalline order. Generally, we find that spin-orbit coupling intertwines magnetic and nematic order, giving rise to ferronematic phases that break translational symmetry. In addition to the usual homogeneous polar and ferromagnetic phases in the non spin-orbit coupled spin-$1$ gas, we find three new phases: a ferromagnetic stripe phase for attractive spin-dependent interactions with stripes in the total density and spin, an XY spiral ferronematic phase with uniform total density for weak repulsive spin-dependent interactions and negative quadratic Zeeman shift, and a biaxial ferronematic stripe phase, with spatial oscillations in the total density, spin vector and nematic director.

A key difference between the spin-$1$ case from the pseudo-spin $1/2$ counterpart is the appearance of spin density wave phases, even when the total density remains uniform for repulsive spin-dependent interactions. We also emphasize that although ferronematic phases are believed to occur in dipolar fermions \cite{erbiumnew, fradkin} and high spin systems such as spin-$3$ Chromium atoms \cite{dienerho}, a crucial difference here is that the ferronematic ground states we find also break translational symmetry, due to the underlying spin-orbit coupling. High spin spin-orbit coupled systems thus offer unique insight into the interplay between competing orders, which are ubiquitous in strongly correlated systems. It will be extremely interesting to explore generalizations of this work to even larger spin systems, such as spin-orbit coupled Dysprosium and Erbium atoms \cite{levdyraman, erbiumnew}. Other exciting directions for future work are to explore the nature of the Goldstone modes associated with broken translation symmetry \cite{stringarigoldstone}, the topological defects that occur in these striped ferronematic phases \cite{zhai, muellerrotating} and the restoration of different symmetries at finite temperature \cite{radic, natuspin1, paramekanti}. 


\textit{Acknowledgements.---} We are grateful to the JQI-NSF-PFC, AFOSR-MURI, and ARO-MURI (Atomtronics) for support.


\begin{thebibliography}{99}
\bibitem{note} In the presence of more exotic Rashba type spin-orbit coupling, the spin density can display spiral order, which spontaneously breaks time-reversal symmetry. (See Refs.~\cite{wang, congjun}) Here time-reveral symmetry is explicitly broken by the Raman coupling term (see Eq.~\ref{soham})
\end{thebibliography}
\end{document}